\documentclass[apj,iop]{emulateapj}
\usepackage{apjfonts, graphicx}

\newcommand{\mycode}[1]{\texttt{#1}} 

\begin{document}

\title{The thermal state of KS~1731$-$260 after 14.5 years in quiescence}
\shortauthors{Merritt et al.}
\shorttitle{Thermal state of KS~1731$-$260}

\author{Rachael~L.~Merritt\altaffilmark{1}}
\author{Edward~M.~Cackett\altaffilmark{1}}
\author{Edward~F.~Brown\altaffilmark{2}}
\author{Dany~Page\altaffilmark{3}}
\author{Andrew~Cumming\altaffilmark{4}}
\author{Nathalie~Degenaar\altaffilmark{5, 6}}
\author{Alex~Deibel\altaffilmark{2}}
\author{Jeroen~Homan\altaffilmark{7}}
\author{Jon~M.~Miller\altaffilmark{8}}
\author{Rudy~Wijnands\altaffilmark{5}}

\email{rachael.merritt@wayne.edu}

\affil{\altaffilmark{1}Department of Physics \& Astronomy, Wayne State University, 666 W. Hancock St, Detroit, MI 48201, USA}
\affil{\altaffilmark{2}Department of Physics \& Astronomy, National Superconducting Cyclotron Laboratory, and the Joint Institute for Nuclear Astrophysics, Michigan State University, East Lansing, MI 48824, USA}
\affil{\altaffilmark{3}Instituto de Astronom\'{i}a, Universidad Nacional Aut\'{o}noma de M\'{e}xico, M\'{e}xico, D. F. 04510}
\affil{\altaffilmark{4}Department of Physics and McGill Space Institute, McGill University, 3600 rue University, Montreal, QC, H3A 2T8, Canada}
\affil{\altaffilmark{5}Anton Pannekoek Institute for Astronomy,  University of Amsterdam, Science Park 904, 1098 XH Amsterdam, the Netherlands}
\affil{\altaffilmark{6}Institute of Astronomy, University of Cambridge, Madingley Road, Cambridge CB3 0HA, United Kingdom}
\affil{\altaffilmark{7}Kavli Institute for Astrophysics and Space Research, Massachusetts Institute of Technology, 70 Vassar Street, Cambridge, MA 02139, USA}
\affil{\altaffilmark{8}Department of Astronomy, University of Michigan, 1085 South University Avenue, Ann Arbor, MI 48109-1104, USA}

\begin{abstract} 
Crustal cooling of accretion-heated neutron stars provides insight into the stellar interior of neutron stars. The neutron star X-ray transient, KS~1731$-$260, was in outburst for 12.5 years before returning to quiescence in 2001. We have monitored the cooling of this source since then through {\it Chandra} and {\it XMM-Newton} observations. Here, we present a 150 ks {\it Chandra} observation of KS~1731$-$260 taken in August 2015, about 14.5 years into quiescence, and 6 years after the previous observation. We find that the neutron star surface temperature is consistent with the previous observation, suggesting that crustal cooling has likely stopped and the crust has reached thermal equilibrium with the core. Using a theoretical crust thermal evolution code, we fit the observed cooling curves and constrain the core temperature (T$_c = 9.35\pm0.25\times10^7$ K), composition (Q$_{imp} = 4.4^{+2.2}_{-0.5}$) and level of extra shallow heating required (Q$_{sh} = 1.36\pm0.18$ MeV/nucleon).  We find that the presence of a low thermal conductivity layer, as expected from nuclear pasta, is not required to fit the cooling curve well, but cannot be excluded either.\end{abstract}

\keywords{stars: neutron --- X-rays: binaries --- X-rays: individual (KS~1731$-$260)}

\section{Introduction}
Transient neutron star low-mass X-ray binaries (LMXBs) provide an opportunity to study the structural properties of neutron stars. Transient LMXBs typically alternate between periods of outburst, i.e. active accretion, and quiescence when accretion rates drop significantly.
During outburst, the system's X-ray emission is dominated by the accretion disk and/or boundary layer..  Accretion onto the neutron star surface can heat the stellar crust out of thermal equilibrium with the core \citep[e.g.][]{ushomirskyrutledge01,rutledge_ks1731_02}. Once the system returns to quiescence, thermal X-ray emission below a few keV originates from the surface of the neutron star \citep{BBR98} and the thermal relaxation of the crust can be observed directly. Direct observation of crustal cooling allows for the extraction of neutron star properties, such as thermal conductivity, crustal structure, and core temperature via cooling models \citep{shternin07,brown09,pagereddy13,turlione15,deibel15,ootes16}.

Crustal cooling of neutron star LMXBs has been observed in eight sources: KS~1731$-$260 \citep{wijnands01,wijnandsetal02,cackett06,cackett10}, MXB~1659$-$29 \citep{wijnands03,wijnandsetal04,cackett06,cackett08,cackett13}, EXO~0748$-$676 \citep{degenaar09,degenaar11a, degenaar14,diaztrigo11}, XTE~J1701-462 \citep{fridriksson10,fridriksson11}, IGR~J17480$-$2446 \citep{degenaar11b,degenaar13,degenaar15}, MAXI~J0556$-$332 \citep{homan14}, Swift~J174805.3$-$244637 \citep{degenaar15}, and potentially Aql X-1 \citep{waterhouse16}. The cooling curves of these sources generally show a significant drop in temperature immediately following their return to quiescence, typically showing a close to exponential decay, with the curves then flattening as the cooling continues \citep[see figure 5 of][for a comparison of several sources]{homan14}. However, the sources show a wide distribution of temperatures at the beginning of quiescence ($\sim$100 -- 300 eV), cooling timescale (e-folding time) and outburst timescales (a few months to nearly 24 years).  While it was initially thought that long (quasi-persistent) outbursts were required to heat the crust out of thermal equilibrium, the observation of crustal cooling in IGR~J17480-2446, Swift~J174805.3$-$244637 and Aql~X-1 demonstrates that it can occur with outbursts of only a few months.  

KS~1731$-$260 was discovered by the \emph{Mir-Kvant} instrument in August 1989 and in previous data it was also observed in outburst in October 1988 \citep{sunyaev89,sunyaev90}. It remained in outburst until early 2001 \citep{wijnands01}. KS~1731-260 was the first source for which crustal cooling was observed and has provided the longest cooling baseline for any source to date. A detailed history of the source can be found in \citet{cackett06}. Since returning to quiescence, monitoring observations with {\it Chandra} and {\it XMM-Newton} have shown continued cooling up until 2009 \citep[8 years into quiescence;][]{wijnands01,wijnandsetal02,cackett06,cackett10}.  Physical models that track the thermal evolution of the neutron star reproduce the observed cooling in KS 1731-260 \citep[e.g.][]{shternin07,brown09,pagereddy13,turlione15,ootes16}.  Fits to the crustal cooling of KS 1731-260 consistently suggest that the crust has a high thermal conductivity indicating a low impurity parameter (i.e. an ordered lattice crustal structure; see below).  Moreover, in order to match the early evolution in the cooling curves it has been found that an additional source of heating at shallow depths in the crust is often needed \citep[e.g.,][]{brown09,degenaar11b, deibel15,waterhouse16}.  Such shallow heating is also required in order to provide the conditions for superburst ignition \citep{cumming06,gasques07}.

Here we present a new \emph{Chandra} observation of KS~1731$-$260, taken six years after the previous \emph{Chandra} observation and a total of approximately 14.5 years into quiescence. In addition to new data, recent updates to the crust cooling model used in \citet{brown09} (see section~\ref{sec:phys}), allow for the exploration of new parameters such as the influence of a nuclear pasta layer and additional shallow heating of the crust. Our findings suggests that cooling has likely halted in KS 1731-260 and that the neutron star crust has returned to thermal equilibrium with the core. In Section 2 we give an overview of data reduction and spectral analysis. In Section 3 we discuss empirical and physical models to fit the quiescent light curve. In Section 4 we discuss our findings and future work.

\section{Data Reduction and Analysis}
The new \textit{Chandra} observation of KS~1731$-$260 was performed over 150 ks in three separate pointings.  A 66 ks segment was performed on 2015 August 6/7 (ObsID: 16734). A 20 ks segment was performed on 2015 August 8 (ObsID: 17706) and 64 ks segment was performed on 2015 August 9 (ObsID: 17707). As with the previous \textit{Chandra} observations, KS~1731$-$260 was at the default aimpoint of the ACIS-S3 chip.  Due to the time since the previous \textit{Chandra} and \textit{XMM-Newton} observations, we decided to reanalyze all data with the latest calibration files and software.  Full details on the previous observations can be found in \citet{wijnands01,wijnandsetal02,cackett06,cackett10}, and also see Table~\ref{tab:specfit}.

\begin{deluxetable*}{lcccccccc}
\tablewidth{0pt}
\tablecolumns{9}
\tablecaption{Neutron Star Atmosphere Fitting Parameters}
\tablehead{
Parameter & 2428 & 013795201/301 & 3796 & 3797 & 0202680101 & 6279/5486 & 10037/10911 & 16734/17706/707\\
& (CXO) & (XMM) & (CXO) & (CXO) & (XMM) & (CXO) & (CXO) & (CXO)}
\startdata
MJD & 51995.1 & 52165.7 & 52681.6 & 52859.5 & 53430.5 &  53512.9 & 54969.7 & 57242.1\\
$kT_{\text{eff}}$$^{\infty}$ (eV) & 104.6 $\pm$ 1.3 & 89.5 $\pm$ 1.03 & 76.4 $\pm$ 1.8 & 73.8 $\pm$ 1.9 & 71.7 $\pm$ 1.4 & 70.3 $\pm$ 1.9 & 64.5 $\pm$ 1.8 & 64.4 $\pm$ 1.2\\
$F_{\text{obs}}$ (10$^{-15}$ ergs cm$^{-2}$ s$^{-1}$) & 41.6 $\pm$ $^{3.4}_{3.6}$  & 16.1 $\pm$ $^{1.10}_{0.97}$ & 5.98 $\pm$ $^{0.71}_{0.91}$ & 4.79 $\pm$ $^{0.58}_{0.63}$ & 3.97 $\pm$ $^{0.53}_{0.51}$ & 3.49 $\pm$ $^{0.61}_{0.41}$ & 1.99 $\pm$ $^{0.30}_{0.36}$ & 1.97 $\pm$ $^{0.23}_{0.22}$\\
$L_{\text{bol}}$ (10$^{33}$ ergs s$^{-1}$) & 2.69 & 1.47 & 0.76 & 0.64 & 0.59 & 0.55 & 0.39 & 0.39
\enddata
\tablecomments{Mass and radius are fixed to 1.4 M$_{\odot}$ and 10 km, respectively. Spectra were modeled using an absorbed neutron star
atmosphere model (nsa) and a photoelectric absorption model (phabs).  
The distance to KS 1731-260 was set to 7 kpc and the column density was tied between observations, giving a best fitting value of
$N_{\rm H}$ = 1.30 $\pm$ 0.06$\times$10$^{22}$ cm$^{-2}$.  
The effective temperature ($kT_{\text{eff}}^{\infty}$) is corrected for gravitational redshift (i.e. it is the effective temperature
for an observer at infinity).  The ObsID and observatory are indicated at the top of the table (CXO = \textit{Chandra} and
XMM=\textit{XMM-Newton}). The observed flux was calculated over the 0.5-10 keV range. The bolometric luminosity was calculated over the $0.01-100$ keV range. We do not include errors on L$_{\rm bol}$ due to large systematic errors.}
\label{tab:specfit}
\end{deluxetable*}

\subsection{Chandra Data Reduction}
We analyzed the \textit{Chandra} data using CIAO (v 4.7) and CALDB (v 4.6.8). Following \citet{cackett06,cackett10} we used a 3\arcsec\ circular extraction region for the source and an annular extraction region for the background with a inner radius of 7\arcsec\ and outer radius of 25\arcsec. The most recent observation had a net count rate of $3.9\pm0.6\times10^{-4}$ counts s$^{-1}$, which is consistent with the previous 2009 \textit{Chandra} observation, which had a net rate of $4.8\pm1.0\times10^{-4}$ counts s$^{-1}$. All observations were reprocessed for the latest calibration files using the \verb|chandra_repro| task. We used \verb|specextract| to extract the spectra and to create the response matrices.

\subsection{XMM Data Reduction}
We analyzed the \textit{XMM-Newton} data using \textit{XMM} Science Analysis Software (SAS) (v 14.0.0).  As with previous analyses, we used a 10\arcsec\ circular extraction region for the source and a 1\arcmin\ circular, off source extraction region for the background.  We reprocessed the observation files using the \verb|emproc| and \verb|epproc| tasks. We used \verb|evselect| to extract the spectra and \verb|rmfgen| and \verb|arfgen| to generate the response matrices.  In both \textit{XMM-Newton} observations there was significant and consistent background flaring. Due to this, we excluded any times when the $>$10 keV light curve had more than 2 counts s$^{-1}$ for MOS 1 and 2 and more than 4 counts s$^{-1}$ for the PN. We also filtered patterns $0-12$ for the MOS and patterns $0-4$ and flag=0 for the PN. The removal of the background flaring eliminated 2\% of the total exposure time for MOS 1 and 2 and 10-15\% of the total exposure time for the PN.

\subsection{Spectral Analysis}
\begin{figure}
\centering
\includegraphics[angle=270, width=\linewidth]{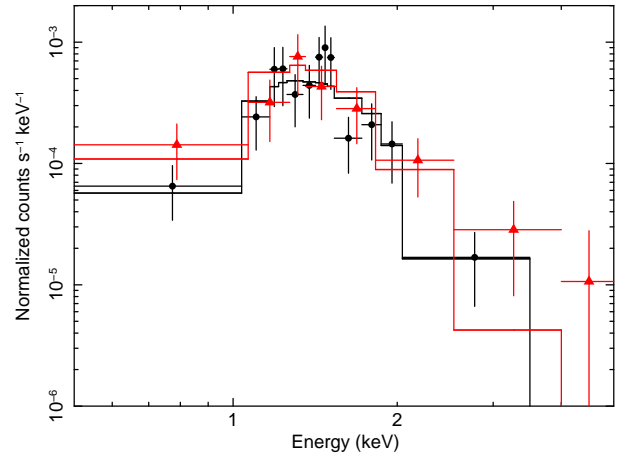}
\caption{The X-ray spectrum of KS~1731$-$260 from the 150 ks {\it Chandra} observation in 2015 August (black, circles) compared to the 2009 May {\it Chandra} observation (red, triangles), along with the best-fitting absorbed neutron star atmosphere modeling.  There is no significant change between 2009 and 2015.  For purposes of clarity in this figure only we have combined separate pointings to create one spectrum for each epoch, and visually rebinned the data.}
\label{fig:spec}
\end{figure}

We fit all available \textit{Chandra} and \textit{XMM-Newton} spectra since the end of outburst using XSPEC \citep[ver. 12.9.0;][]{arnaud96} following a procedure similar to \citet{cackett06,cackett10}.  We modeled the spectra using a neutron star atmosphere model \citep[nsa;][]{zavlinetal96}, modified by photoelectric absorption within our Galaxy (phabs). The nsa model has been used previously in studies of KS~1731$-$260, and other models, such as nsatmos \citep{heinke06}, provide consistent results within 1$\sigma$. A comparison of atmosphere models can be found in Section 3.4 of \citet{heinke06}. For our analysis and modeling, we fix the neutron star radius to canonical values of 10~km and the mass to 1.4~M$_{\odot}$. We set the distance to the source at 7 kpc \citep{muno00}, resulting in a normalization parameter = $2.041\times10^{-8}$ pc$^{-2}$.  The normalization is also a fixed parameter.  We show the X-ray spectrum, and best-fitting model, for the new {\it Chandra} observation in Figure~\ref{fig:spec}, compared to the 2009 {\it Chandra} observation.  The latest spectrum is still thermal, with no need for an additional power-law component, and shows no significant change between 2009 and 2015 (see Figure~\ref{fig:spec}).

All spectra were fit simultaneously with absorption column density, $N_{\rm H}$, tied between all observations and the effective temperature set as a free parameter.  Due to the close proximity of some observations, several spectra had their parameters tied together.  We tied the parameters of the spectra for \textit{XMM-Newton} observations taken around MJD 52165.7 (ObsIDs 012795201/301) and \textit{Chandra} observations taken around MJD 53512.9 (ObsIDs 6279/5468), MJD 54969.7 (ObsIDs 10037/10911), and MJD 57242.1 (ObsIDs 16734/17706/17707).  Using GRPPHA, we binned the spectra to have one count per bin.  We used C-statistic to fit the binned spectra.  C-statistic was used rather than $\chi^2$ statistics due to low number of total counts in some spectra.  

\begin{figure}
\centering
\includegraphics[angle=270,width=\linewidth]{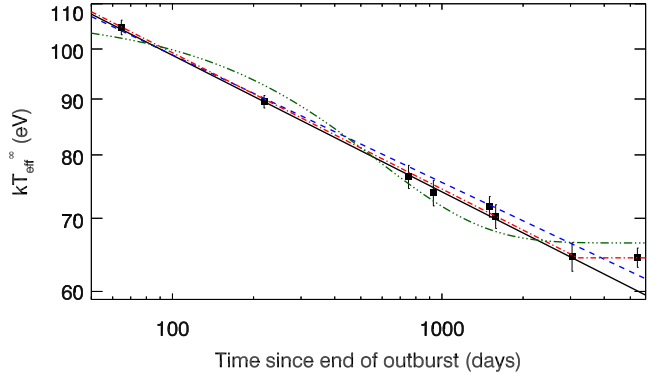}
\caption{Effective temperature for KS~1731$-$260 over approximately 5300 days since the end of outburst.  The lines indicate empirical model fits for a power-law fit to the first 7 data points, i.e. excluding the most recent one (black, solid line); a power-law fit to all 8 data points (blue, dashed line); an exponential decay to a constant (green, dash-dot-dot-dot line); a broken power-law (red, dash-dot line).}
\label{fig:bestmodel}
\end{figure}

The spectral fitting results are shown in Table~\ref{tab:specfit}. The effective temperature is corrected for gravitational redshift (i.e. it is the effective temperature for an observer at infinity).  The new {\it Chandra} observation gives $kT_{\text{eff}}^{\infty} = 64.4 \pm1.2$~eV, this temperature is consistent with the previous 2009 \textit{Chandra} observation ($kT_{\text{eff}}^{\infty} = 64.5 \pm 1.8$~eV) within 1$\sigma$, suggesting that the crust of KS~1731$-$260 has returned to thermal equilibirum with the core.  The progression of effective temperature over time (cooling curve) is shown in Figure~\ref{fig:bestmodel}.

\section{Cooling Curves}

\subsection{Empirical Models}
Prior to the 2015 observation, the cooling curve of KS~1731$-$260 was well fit by a simple power-law \citep{cackett08,cackett10} or an exponential decay to a constant \citep{cackett06,cackett10}.  We fit both models to all data points (see Figure~\ref{fig:bestmodel}).  The exponential decay to a constant  (green dash-dot-dot-dot line in Fig.~\ref{fig:bestmodel}), provides a poor fit with $\chi^2 = 20.3$ for 5 degrees of freedom (d.o.f.), while the power-law provides a significantly better fit (blue dashed line in Fig.~\ref{fig:bestmodel}), with $\chi^2 = 7.76$ for 6 d.o.f.    We used a power-law of the form $y(t)=\alpha(t-t_0)^\beta$, where $t_0$ corresponds to midday of the last observation of KS~1731$-$260 in outburst \citep[MJD 51930.5,][]{cackett06}.  The best fitting parameters are $\alpha =169.0\pm0.9$~ eV, and $\beta = -0.117\pm0.004$~eV/day.  This power-law fit bisects the last two points, given that the last point has the same temperature as the previous one.  We therefore also try a broken power-law model, of the form  $y(t)=\alpha \left( \frac{t-t_0}{t_{\rm br}} \right)^\beta$ for $t < t_{\rm br}$ and $y(t) = \alpha$ for $t \geq t_{\rm br}$ (red, dash-dot line in Fig.~\ref{fig:bestmodel}).  This provides a better fit, with $\chi^2 = 1.26$ for 5 d.o.f., with best fitting parameters of $\alpha = 64.4\pm1.2$~eV, $\beta = -0.125\pm0.005$~eV/day and $t_{\rm br} = 3200\pm600$ days.  Using an F-test, this improvement in $\chi^2$ is significant at the 2.9$\sigma$ confidence level.

Given suggestions that cooling has stopped, we also fit a power-law to just the first 7 observations, i.e., excluding the latest one (see the black solid line in Fig~\ref{fig:bestmodel}) and get best fitting parameters $\alpha=175.5 \pm 2.9$~eV and $\beta= -0.125 \pm 0.002$~eV/day.  Extrapolating this power-law to the time of the latest observation gives $kT_{\text{eff}}$$^{\infty}$ = 60.2 eV. The measured effective temperature for the newest observation ($kT_{\text{eff}}$$^{\infty}$ = 64.4 $\pm$ 1.2 eV) therefore deviates from the extrapolation of the previous cooling behavior at the 3.5$\sigma$ level. This further suggests that the crust of KS~1731$-$260 may have stopped cooling.

\subsection{Physical Models}\label{sec:phys}
In order to fit the quiescent crustal cooling, we model the thermal evolution of the neutron star using the open-source code \mycode{dStar}\footnote{\url{https://github.com/nworbde/dStar}} \citep{brown15}.  This code uses the same microphysics and integration scheme discussed in \citet{brown09}.  In addition, the code now has a more flexible interface that allows the distribution of heat sources and impurities with depth specified by the user; this allows us to model the effect of an additional heat source in the shallow ocean and an insulating pasta layer in the deep crust. Fixed parameters in \mycode{dStar} include the crust composition, where the composition of \citet{haensel90} is used, the atmosphere model with the column depth of the light element layer set to $10^4\,\mathrm{g\, cm^{-2}}$, crust-core transition density of 8.13$\times10^{13}$ g cm$^{-3}$, and the radius and mass are set to 10~km and 1.4~M$_\odot$ to be self-consistent with the spectral fits.  Our \mycode{dStar} models use the superfluid critical temperature in the crust from \citet{schwenk03}.

\begin{figure}
\centering
\includegraphics[angle=270,width=0.9\linewidth]{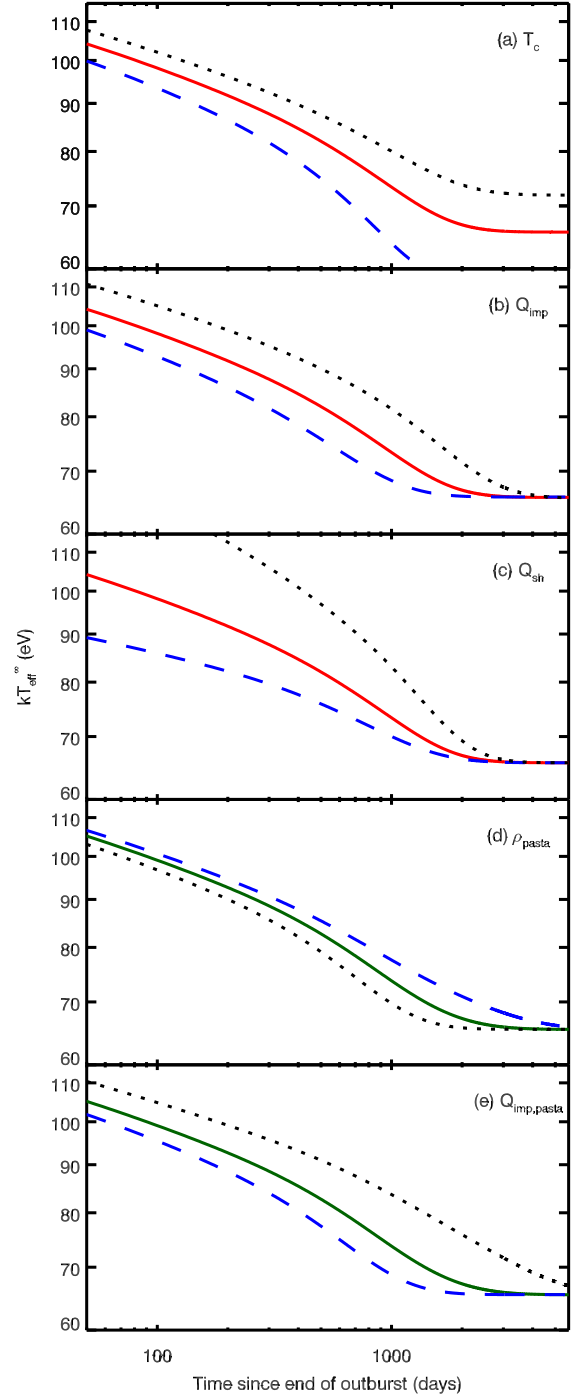}
\caption{Influence of physical parameters in \mycode{dStar}. In panels $(a) - (c)$, the solid red line is the best fit model without a nuclear pasta layer. In panels $(d) - (e)$, the solid green line is the best fit model with nuclear pasta. In each panel we change only one parameter, while leaving the others at their best fit values. {\it (a)} the influence of changing the core temperature with dashed blue line showing T$_c=7.35\times10^7$~K and dotted black line showing T$_c=1.1\times10^8$~K. {\it (b)}, the influence of changing the impurity parameter, with dashed blue line showing Q$_{imp} = 1.0$ and dotted black line showing Q$_{imp} = 10.0$. {\it (c)} The influence of changing the amount of shallow heating with the dashed blue line showing Q$_{sh}=0.60$~MeV/nucleon, and the dotted black line showing Q$_{sh} =3.5$~MeV/nucleon. {\it (d)}, The influence of changing the transition density of the pasta with the dashed blue line showing $\rho_{\rm pasta}=1\times10^{13}$~g~cm$^{-3}$, and the dotted black line showing $\rho_{\rm pasta}=6\times10^{13}$~g~cm$^{-3}$. {\it (e)} The influence of changing the impurity parameter of the pasta, with the dashed blue line showing Q$_{imp,pasta}=2.0$ and the dotted black line showing Q$_{imp,pasta}=40$.}
\label{fig:parvar}
\end{figure}

Prior to modeling the cooling, we simulate 12.5 years of constant accretion at the rate of $\dot{M} = 10^{17}$ g s$^{-1}$ which is consistent with the time averaged accretion rate found by \citet{galloway08}.  For the initial run of models we varied three parameters: core temperature (T$_c$), the impurity parameter (Q$_{imp}$$\equiv$ n${^{-1}_{ion}}$$\sum_{i}$n$_i$(Z$_i$ - $\langle$Z$\rangle$)$^2$) of the crust, and additional shallow heating of the crust (Q$_{sh}$). The impurity parameter measures the distribution of nuclide charge numbers.  This influences the conductivity and structure of the crust  (i.e. low Q$_{imp}$ means high thermal conductivity and well structured lattice).  The shallow heating is an additional few MeVs of crustal heating caused by an unknown source \citep[e.g.][]{cumming06,brown09,medin14, deibel15}. In the model, we set the depth of the shallow heating to occur between pressure values of P$_{Q_{sh},min} = 1\times10^{27}$ g cm$^{-1}$ s$^{-2}$ and  P$_{Q_{sh},max} = 1\times10^{28}$ g cm$^{-1}$ s$^{-2}$.  The pressure values were chosen so that the heat source was shallower than the depth corresponding to the thermal time of the first observation, but otherwise it was arbitrary.  The cooling curve is not that sensitive to these boundaries since the first observation is 65 days after the end of the outburst. The influence of the $T_c$, Q$_{imp}$ and $Q_{sh}$ parameters on the shape cooling curve can be seen in panels (a) -- (c) of Figure~\ref{fig:parvar}.

\begin{figure}
\centering
\includegraphics[angle=270, width=\linewidth]{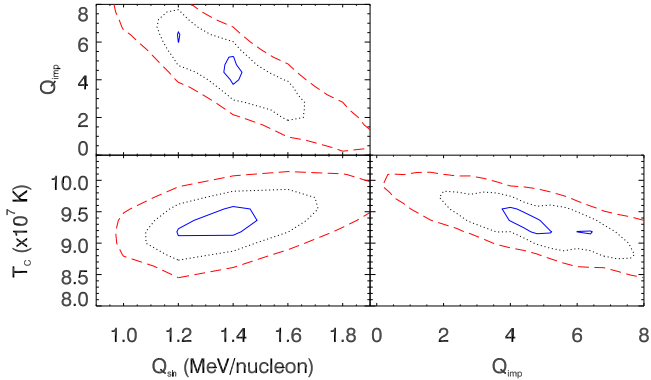}
\caption{Contour plots of $\chi$$^2$ distributions of the parameter space of the models without nuclear pasta. 1$\sigma$ is designated by the solid blue line, 2$\sigma$ by the dotted black line, and 3$\sigma$ by the dashed red line. }
\label{fig:nopastacontours}
\end{figure}

\begin{figure}
\centering
\includegraphics[angle=270,width=\linewidth]{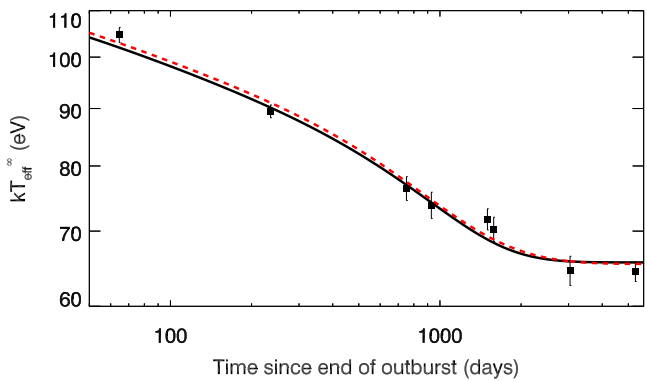}
\caption{Comparison of best-fitting physical models from \mycode{dStar}. The solid black line is the best fit model without nuclear pasta., while the dashed red line is the best fit model including a nuclear pasta layer. Both cooling curves (with and without pasta) have similar shapes.}
\label{fig:physmod}
\end{figure}

We calculate the thermal evolution of the crust for a wide range in all parameters, creating a grid of cooling curve models.  This is the first time a full exploration of parameter space has been conducted. To find the best fit parameters, we search the grid to find the model which fits the data with the lowest $\chi^2$ value.  The best fitting parameters are T$_c = 9.35\pm0.25\times10^7$ K,  Q$_{imp} = 4.4^{+2.2}_{-0.5}$, and Q$_{sh} = 1.36\pm0.18$ MeV/nucleon. The uncertainties quoted here, and throughout the paper are at the 1$\sigma$ level. The best fitting model is shown as a solid black line in Figure~\ref{fig:physmod}.  This model gives a reduced-$\chi$$^2$ = 2.00 (d.o.f. = 5). Our core temperature value is approximately 2 times greater than the core temperature values found in \citet{cackett10} (T$_c = 4.6\times10^7$ K) and \citet{brown09} (T$_c = 5.4\times10^7$). We find our impurity parameter value is higher than but similar to \citep{cackett10} (Q$_{imp}$ = 4.0) and approximately three times greater than \citet{brown09}  (Q$_{imp}$ = 1.5). The slightly different core temperature is not surprising since it is strongly constrained by this latest data point, since it is the surface temperature once the crust and core are in thermal equilibrium.  Moreover, our choice of light-element column depth is smaller than in \citet{brown09} (who use $10^9\,\mathrm{g\, cm^{-2}}$), the effect of which leads to a higher core temperature because the envelope is less opaque \citep{cumming16}. The 1, 2 and 3$\sigma$ confidence regions of the parameter space can be seen in Figure~\ref{fig:nopastacontours}.

\begin{figure*}
\centering
\includegraphics[angle=270, width=\linewidth]{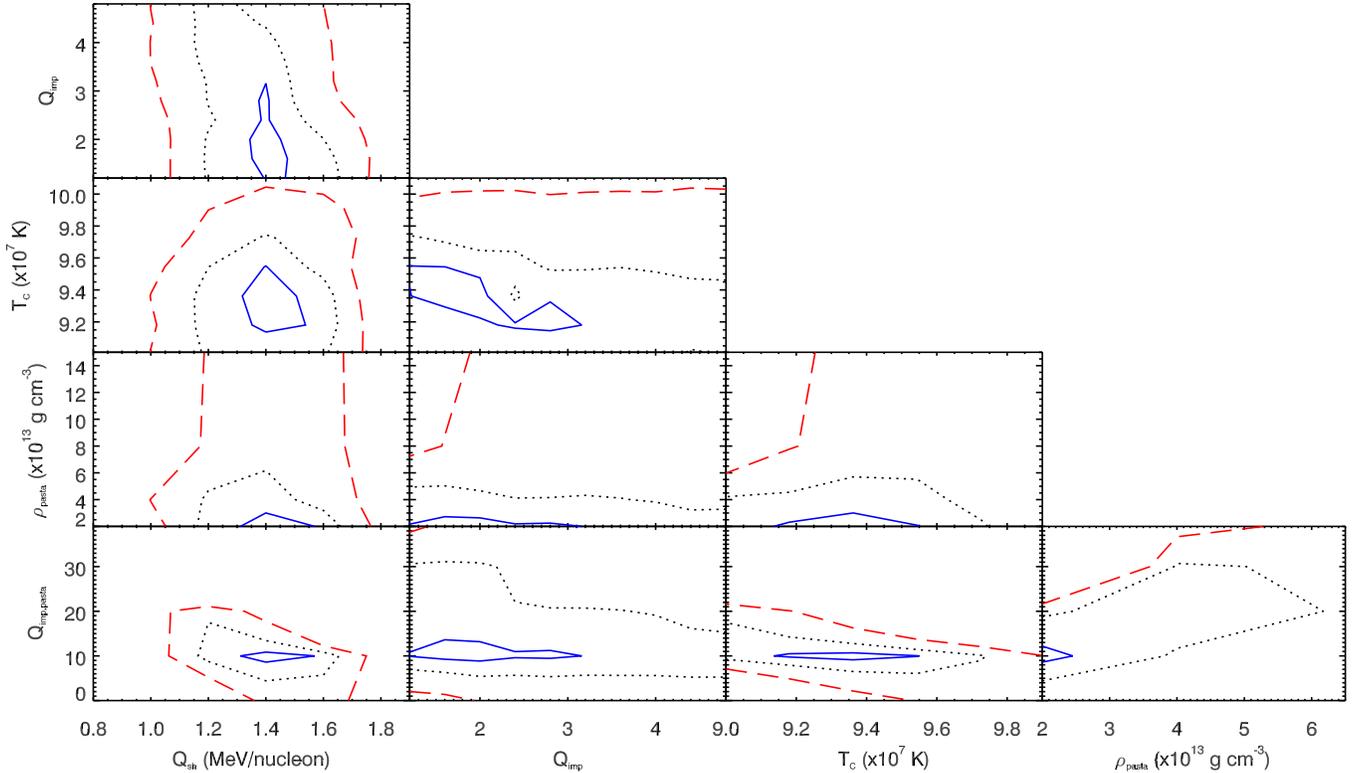}
\caption{Contour plots of $\chi$$^2$ distributions of the parameter space of the nuclear pasta models. 1$\sigma$ is designated by the solid blue line, 2$\sigma$ by the dotted black line, and 3$\sigma$ by the dashed red line. }
\label{fig:pastacontours}
\end{figure*}

We ran a second set of physical models that included a low thermal conductivity layer, which is consistent with nuclear pasta.  A nuclear pasta phase is expected to exist at the base of neutron star crusts \citep[e.g., see][and references therein]{horowitz15}.  \citet{pons13} suggested that the irregular shapes in the pasta would lead to low conductivity.  \citet{horowitz15} showed that there could be defects in the pasta and suggested that they could act as scattering sites that would lower the thermal conductivity. We kept T$_c$, Q$_{imp}$, and Q$_{sh}$ of the crust as free parameters. We added Q$_{imp}$ of the pasta layer and the density of the transition to the pasta phase, $\rho_{\rm pasta}$, as variable parameters.  The pasta layer is insulating. The presence of a pasta phase requires a greater temperature gradient to carry a thermal flux.  As the transition density becomes smaller, the pasta layer becomes thicker, which causes the temperature of the crust to increase and the crust cools more slowly.  Additionally, the pasta layer becomes more insulating as the impurity parameter is increased, resulting in a similar effect as the decreased transition density. The influence of the pasta parameters on the cooling curve can be seen in panels (d) -- (e) of Figure~\ref{fig:parvar}.  Again, we create a large grid of models, and find the model with the best (lowest) $\chi^2$ value. The best fitting parameters are T$_c = 9.34\pm0.21\times10^7$ K,  Q$_{imp} = 2.1\pm1.0$, Q$_{sh} = 1.43\pm0.15$ MeV/nucleon, Q$_{imp,pasta} = 12.4\pm5.1$, and $\rho_{pasta} = 2.7\pm0.8\times10^{13}$ g cm$^{-3}$. The best-fitting pasta model is shown as a dashed red line in Figure~\ref{fig:physmod}.  This model gives a reduced-$\chi$$^2$ = 3.232 (dof=3). The 1, 2 and 3 $\sigma$ confidence regions of the parameter space can be seen in Figure~\ref{fig:pastacontours}.

\section{Discussion}
We have analyzed the newest \emph{Chandra} observation of the neutron star LMXB KS~1731$-$260, 14.5 years into quiescence and six years after the previous \emph{Chandra} observation.  From fitting the X-ray spectrum we find an effective temperature of $kT_{\text{eff}}$$^{\infty}$ = 64.4 $\pm$ 1.2 eV, which is consistent (within 1$\sigma$) of the 2009 \emph{Chandra} observation which had $kT_{\text{eff}}$$^{\infty}$ = 64.5 $\pm$ 1.8 eV, implying that the crust has likely thermally relaxed, and may have returned to equilibrium with the core.  A broken power-law fits the cooling curve better than a simple power-law, at approximately the 3$\sigma$ level.  Alternatively, if we exclude the latest data point and re-fit a simple power-law decay we find that the newest observation is 3.5$\sigma$ away from an extrapolation of the best-fitting model, further suggesting the crust has stopped cooling.  We caution, however, that further observations are needed to confirm that this is not a deviation from the cooling curve caused by low-level accretion, as has been seen in, e.g., XTE~J1701$-$462 \citep{fridriksson11}.  The lack of a power-law component in this latest {\it Chandra} observation of KS~1731$-$260 suggests this may not be happening here.

We constrain crust properties using a physical model of the crust's thermal relaxation.  This model calculates the thermal evolution of crust, and we initially allow the core temperature, impurity parameter and presence of additional shallow heat sources in the crust as free parameters to fit the cooling curve.  We found best fit values for the core temperature and impurity parameter (T$_c = 9.35\pm0.25\times10^7$  K and  Q$_{imp} = 4.4^{+2.2}_{-0.5}$). The low impurity parameter value is consistent with previous discussions that the crust of KS~1731$-$260 has high thermal conductivity \citep[e.g.,][]{wijnandsetal02,cackett06,shternin07,brown09,pagereddy13}.  The core temperature of the neutron star is most sensitive to cooling at late times. This new late time observation ($\sim$5300 days after the end of outburst), along with the full exploration of parameter space, places significantly improved constraints on the core temperature. In turn, constraints on the core temperature are important for placing a lower limit on the core's heat capacity \citep{cumming16}.  Moreover, we find that significant additional shallow heat source in the crust is preferred, with a best-fitting value of Q$_{sh} = 1.36\pm0.18$ MeV/nucleon.  This amount of extra heating is consistent with that found in other neutron star LMXBs \citep[e.g., EXO~0748$-$676 and Aql~X-1;][]{degenaar14,waterhouse16}, but is significantly less than the amount of shallow heating required for MAXI~J0556$-$332 \citep[Q$_{sh} \approx 4-10$ MeV/nucleon;][]{deibel15}.

\mycode{dStar} also allows to test for the presence of a layer of low thermal conductivity material close to the crust-core transition, possibly a nuclear pasta phase. However, we find that the inclusion of such a layer did not have significant influence on the best-fitting model, with the best fit coming from the lowest pasta density we allowed from the model (see contours in Figure~\ref{fig:pastacontours}).  This is consistent with \citet{horowitz15} that KS~1731$-$260 can be modeled equally as well with or without a nuclear pasta layer. A parameter we did not vary in our pasta model was the superfluid critical temperature in the crust. The influence of this parameter can be seen in \citet{deibel16}.

Our modeling with \mycode{dStar} assumes a constant accretion rate throughout the 12.5 year outburst of KS~1731$-$260.  However, a recent investigation by \citet{ootes16} has shown that variations in the accretion rate throughout the outburst influences the cooling curve.  Especially important are variations at the end of the outburst, which can strongly influence the early part of the cooling curve, and hence have a significant impact on the implied amount of extra heating at shallow depths.  Their modeling of KS~1731$-$260 (without including the latest observation) finds the need for a 1.4 MeV/nucleon shallow heat source when variations in accretion rate are included and only 0.6 MeV/nucleon when a constant accretion rate is assumed (though, note, that \citet{ootes16} do not optimize their fits to get the best-fit parameters).  This value for Q$_{sh}$ is significantly less than the value implied from our modeling, suggesting some dependence on the input parameters that we do not fit for, such as the depth of the shallow heating and crust composition included in the cooling code.  For instance, the minimum shallow heating density we use is $\rho_{sh,min} = 2.12\times10^{9}$ g cm$^{-3}$, while \citet{ootes16} use $\rho_{sh,min} = 4\times10^{8}$ g cm$^{-3}$.  The choice of $\rho_{sh,min}$ was used in previous calculations and we have tested that our results are insensitive to the precise value.
  
KS~1731$-$260 has shown superbursts \citep{kuulkers02}, type-I X-ray bursts due to thermonuclear burning of carbon, leading to bursts that last for hours rather than tens of seconds.  The conditions required for carbon to ignite places constraints on the thermal properties of the crust \citep{cumming06}, independently to the constraints placed by crustal cooling.  The presence of superbursts requires that the crust has a temperature of $\sim6\times10^8$~K at a depth of $\sim10^{12}$ g cm$^{-2}$ \citep{cumming06}.  Without any additional heat source the crust is too cold \citep{cumming06,gasques07}.  \mycode{dStar} gives the crust temperature profile at all times, thus we can study the temperature in the crust at the time of the superburst.  As discussed in \citet{ootes16}, the accretion rate during the outburst varies, and is significantly higher in the earlier part than at the end.  The persistent flux before the superburst \citep[taken from][]{kuulkers02} is about 1.66 times higher than the average mass accretion rate we assume.  We therefore recalculate the thermal evolution of the crust during the outburst assuming this higher mass accretion rate, and a correspondingly higher value of $Q_{sh}$, and look at the crust temperature profile at a time of 2894 days into the outburst (when the superburst took place).  Doing  this gives a temperature in the outer part of the crust (close to $10^{12}$ g cm$^{-2}$) of $\sim5\times10^8$~K.  Which, is broadly consistent with the required $6\times10^8$~K for superburst ignition, within the uncertainties and assumptions made.

KS~1731$-$260 provides the rare opportunity to place two independent mass-radius constraints on the same source.  During outburst, KS~1731$-$260 displayed photospheric radius expansion (PRE) bursts \citep{muno00}. PRE bursts are thought to reach the Eddington luminosity, thus fitting the blackbody emission from the burst can lead to both mass and radius constraints (both the Eddington luminosity and observed emitted radius depend on M and R), assuming the distance to the source is known (\citealt{ozel06}, though see e.g. \citealt{poutanen14} for an opposing view). \citet{ozel12} applied this technique to KS~1731$-$260, implying $R \leq 12.5$ km and $M  \leq 2.1$ M$_\odot$ (95\% confidence level).  In future work, we will explore constraints on M and R through fitting crustal cooling models to KS~1731$-$260.

\acknowledgements
RLM and EMC gratefully acknowledge support for this work provided by the National Aeronautics and Space Administration through Chandra Award Number GO5-16043X issued by the Chandra X-ray Observatory Center, which is operated by the Smithsonian Astrophysical Observatory for and on behalf of the National Aeronautics Space Administration under contract NAS8-03060.  EFB and AD are supported by the National Science Foundation under Grant No. AST-1516969.  AC is supported by an NSERC Discovery Grant. ND is supported by an NWO/Vidi grant and an EU Marie Curie Intra-European fellowship (contract no. FP-PEOPLE-2013-IEF-627148). RW acknowledges support from a NWO top grant, module 1. DP is partially supported by the Consejo Nacional de Ciencia y Tecnolog{\'\i}a with a CB-2014-1 grant $\#$240512.  This work was enabled in part by the National Science Foundation under Grant No. PHY-1430152 (JINA Center for the Evolution of the Elements).  We also thank the International Space Science Institute, Bern, for hosting the Neutron Star Crust Team where many productive discussions on this topic took place.

\bibliographystyle{apj}
\bibliography{apj-jour,qNS}

\end{document}